\documentclass[12pt,a4paper]{article}

\usepackage{epsfig}
\usepackage{psfig}
\def\lsim{\raise0.3ex\hbox{$<$\kern-0.75em\raise-1.1ex\hbox{$\sim$}}}
\def\gsim{\raise0.3ex\hbox{$>$\kern-0.75em\raise-1.1ex\hbox{$\sim$}}}
\newcommand{\be}{\begin{equation}}
\newcommand{\ee}{\end{equation}}
\newcommand{\ba}{\begin{eqnarray}}
\newcommand{\ea}{\end{eqnarray}}

\def\spose#1{\hbox to 0pt{#1\hss}}
\def\ltapprox{\mathrel{\spose{\lower 3pt\hbox{$\mathchar"218$}}
 \raise 2.0pt\hbox{$\mathchar"13C$}}}
\def\gtapprox{\mathrel{\spose{\lower 3pt\hbox{$\mathchar"218$}}
 \raise 2.0pt\hbox{$\mathchar"13E$}}}

\def\NT{N_\tau}
\def\nt{\ifmmode\NT\else$\NT$\fi}
\def\NS{N_\sigma}
\def\ns{\ifmmode\NS\else$\NS$\fi}

\def\n{\noindent}

\begin{document}
\begin{titlepage} 
\thispagestyle{empty}

 \mbox{} \hfill BI-TP 99/35~\\
 \mbox{} \hfill November 1999
\begin{center}
\vspace*{1.0cm}
{{\Large \bf Goldstone-mode effects and scaling function \\         
  for the three-dimensional $O(4)$ model\\}}\vspace*{0.5cm}
{\large J\"urgen Engels and 
Tereza Mendes}\\ \vspace*{1.0cm}
\centerline {{\em Fakult\"at f\"ur Physik, Universit\"at Bielefeld, 
D-33615 Bielefeld, Germany}} \vspace*{0.5cm}
\protect\date \\ \vspace*{0.9cm}
{\bf   Abstract   \\ } \end{center} \indent
We investigate numerically the three-dimensional $O(4)$ model 
on $24^3$--$120^3$ lattices as a function of the magnetic field $H$.
We verify explicitly the singularities induced by Goldstone modes in the
low-temperature phase of the model, and show that they are also 
observed close to the critical temperature. Our results are well
described by the perturbative form of the model's magnetic equation 
of state, with coefficients determined nonperturbatively from our data.
The resulting expression is used to generate the magnetization's
scaling function parametrically.
\vfill \begin{flushleft} 
\noindent{\rule[-.3cm]{5cm}{.02cm}} \\
\vspace*{0.2cm} 
E-mail: engels@physik.uni-bielefeld.de,
        mendes@physik.uni-bielefeld.de. \end{flushleft} 
\end{titlepage}


\section{Introduction}

The $O(N)$ spin models (or, more precisely, the $O(N)$-invariant
nonlinear $\sigma$-models) are defined by
\be
\beta\,{\cal H}\;=\;-J \,\sum_{<i,j>} {\bf S}_i\cdot {\bf S}_j
         \;-\; {\bf H}\cdot\,\sum_{i} {\bf S}_i \;,
\ee
where $i$ and $j$ are nearest-neighbour sites on a $d-$dimensional 
hypercubic lattice, and ${\bf S}_i$ is an $N$-component unit vector 
at site $i$. The case $N=1$ corresponds to the Ising model. 
We will consider here only $N > 1$. It is convenient to decompose 
the spin vector ${\bf S}_i$ into longitudinal (parallel to the magnetic 
field ${\bf H}$) and transverse components 
\be
{\bf S}_i\; =\; S_i^{\parallel} {\bf \hat H} + {\bf S}_i^{\perp} ~.
\ee
The order parameter of the system, the magnetization $M$, is then the 
expectation value of the lattice average $S^{\parallel}$
of the longitudinal spin components
\be
M \;=\; <\!\frac{1}{V}\sum_{i} S_i^{\parallel}>\; =\; <  S^{\parallel}>~.
\ee
Two types of susceptibilities are defined. The longitudinal 
susceptibility is the usual derivative of the magnetization, 
whereas the transverse susceptibility corresponds to the fluctuation 
per component of the lattice average ${\bf S}^{\perp}$ of the
transverse spin components
\ba
\chi_L\!\! &\!=\!&\!\! {\partial M \over \partial H}
 \;=\; V(<{ S^{\parallel}}^2>-M^2)~, \label{chil}\\
\chi_T\!\! &\!=\!&\!\! V{1 \over N-1} < {{\bf S}^{\perp}}^2> 
\;=\;{M \over H}~. \label{chit}
\ea

These models are of general interest in condensed matter 
physics, but have applications also in quantum field theory.
In particular, the three-dimensional $O(4)$ model is of importance for 
quantum chromodynamics (QCD) with two degenerate light-quark flavors 
at finite temperature. If QCD undergoes a second-order chiral transition
in the continuum limit, it is believed to belong to the same universality 
class as the $3d$ $O(4)$ model \cite{PW,RW}. QCD lattice data have therefore
been compared to the $O(4)$ scaling function, determined numerically in
\cite{Toussaint}. For staggered fermions this comparison is at present
not conclusive \cite{MILC}, but results for Wilson fermions 
\cite{wilson} seem to agree quite well with the predictions.

$O(N)$ models in dimension $2<d\leq4$
are predicted to display singularities on the coexistence line 
$T<T_c, H=0$ due to the presence of massless 
Goldstone modes \cite{goldstone}. 
In fact, both susceptibilities are predicted to diverge 
in this region.
The magnetic equation of state is nevertheless
divergence-free, and compatible with these singularities.
The equation of state was calculated up to order 
$\epsilon^2$ in the $\epsilon$-expansion by Brezin et al.\ \cite{Brezin}.
On the basis of this expansion it has been argued \cite{WZ} that these 
singularities may be easily observable, since the perturbative
coefficient associated with the diverging term is quite large.

A further consequence of the Goldstone singularities is the 
appearance of strong finite-size effects at all $T<T_c$ for $H\to0$.
These effects have been studied using chiral perturbation theory
in \cite{clust,cpt}. 
Direct numerical evidence of the 
Goldstone singularities is however lacking, apart 
from early simulations of the three-dimensional $O(3)$ model 
on small lattices \cite{MK}, where indications of the 
predicted behaviour were found.

The aim of this paper is to verify explicitly the 
Goldstone singularities, and to investigate their interplay
with the critical behaviour and the effect they have on the scaling 
function. We do this by simulating the 
three-dimensional $O(4)$ model in the presence of an external 
magnetic field in the low-temperature phase and close to the critical
temperature $T_c$. 
For determining the scaling function, we have also simulated at some 
high-temperature values. 
First results of our work have been presented at {\em Lattice'99} 
\cite{lat99}. 
The plan of the paper is as follows.
In the next section we review the perturbative predictions for the 
magnetization and the susceptibilities at low temperatures, 
as well as the analytic results for the magnetic equation of state,
which is equivalent to the magnetization's scaling function.
Our numerical results are discussed in Section \ref{section:results}. 
The fits and parametrization for the scaling function are given
in Section \ref{section:sca_fun}. 
A summary and our conclusions are presented in Section
\ref{section:conclusion}.


\section{$\!\!$Perturbative Predictions and Critical Behaviour}
\label{section:PT}


The continuous symmetry present in the $O(N)$ spin models
gives rise to the so-called {\em spin waves}: slowly varying
(long-wavelength) spin configurations, whose energies may be arbitrarily
close to the ground-state energy. In two dimensions these modes are
responsible for the absence of spontaneous magnetization,
whereas in $d>2$ they are the massless Goldstone modes associated 
with the spontaneous breaking of the rotational symmetry for
temperatures below the critical value $T_c$ \cite{spin}.
For $T<T_c$ the system is in a broken phase, i.e.\ the
magnetization $M(T,H)$ attains a finite value $M(T,0)$ at $H=0$.
To be definite we assume here $H>0$. 
As a consequence the transverse susceptibility, which is 
directly related to the fluctuation of the Goldstone modes, 
diverges as $H^{-1}$ when $H\to0$ for all $T<T_c$. 
This can be seen immediately from the identity
\be
\chi_T \;=\; \frac{M(T,H)}{H}~.
\label{WI}
\ee
This expression is a direct 
consequence of the $O(N)$ invariance of the zero-field free energy, 
and can be derived as a Ward identity \cite{Brezin}. It is valid for
all values of $T$ and $H$.

A less trivial result \cite{goldstone,WZ} is that also the longitudinal 
susceptibility is diverging on the coexistence curve for 
$2<d\leq4$. The leading term in the perturbative expansion
for $2<d<4$ is $H^{d/2-2}$. The predicted divergence 
in $d=3$ is thus
\be
\chi_L(T<T_c,H)\;\sim\; H^{-1/2}~.
\label{chiL}
\ee
This is equivalent to an $H^{1/2}$-behaviour of the magnetization near 
the coexistence curve 
\be
M(T<T_c,H)\;=\;M(T,0)\,+\,c\,H^{1/2}~.
\label{magn}
\ee

An interesting question is whether the above expressions
still describe the behaviour close to the critical region $T\ltapprox T_c$.
We recall that the critical behaviour is determined by the singular 
part of the free energy.
Its scaling form in the thermodynamic limit is  
\be
f_s(t,h)\;=\;b^{-d}f_s(b^{y_t} t, b^{y_h} h)~.
\ee
Here we have neglected possible dependencies on irrelevant scaling fields
and exponents. The variables $t$ and $h$ are the conveniently normalized 
reduced temperature $t=(T-T_c)/T_0$ and magnetic field $h=H/H_0$, and $b$ 
is a free length rescaling factor. The relevant exponents $y_{t,h}$ specify
all the other critical exponents
\be
y_t=1/\nu, \quad y_h=1/\nu_c~;
\ee
\be
\nu_c=\nu/\beta\delta,\quad d\nu=\beta(1+\delta),\quad\gamma=\beta(\delta-1)~.
\ee
Choosing the scale factor $b$ such that $b^{y_h} h=1$ and using 
$M=-\partial f_s/\partial H$ one finds the equation
\be
M\;=\;h^{1/\delta} f_G(t/h^{1/\beta\delta})~,
\label{ftous}
\ee
where $f_G$ is a scaling function. It becomes universal after fixing
the normalization constants $H_0$ and $T_0$. This scaling function 
was calculated numerically for the $3d$ $O(4)$ model by Toussaint 
\cite{Toussaint} and is used in comparison to QCD lattice data
\cite{MILC,wilson}.

Alternatively, one may choose~$b^{y_t}|t|=1$. This leads to the 
Widom-Griffiths form of the equation of state \cite{Griffiths}
\be
y\;=\;f(x)\;,
\label{eqstate}
\ee
where 
\be
y \equiv h/M^{\delta}, \quad x \equiv t/M^{1/\beta}.
\label{xy}
\ee
It is usual to normalize $t$ and $h$ such that
\be
f(0) = 1, \quad f(-1) = 0~.
\label{normal}
\ee
The scaling forms in Eqs.\ (\ref{ftous}) and (\ref{eqstate}) are 
clearly equivalent. In the following we will work with form 
(\ref{eqstate}) and obtain (\ref{ftous}) from it parametrically
in Section \ref{section:sca_fun}.

The equation of state (\ref{eqstate}) has been derived by Br\'ezin
et al.\ \cite{Brezin} to order $\epsilon^2$ in the 
$\epsilon$-expansion, where $\epsilon=4-d$. 
Although diverging terms in $\chi_T$ appear at intermediate steps 
of the derivation, they are canceled by diverging $\chi_L$ terms, 
and the resulting expression is divergence-free.
This expression has been considered by Wallace and Zia \cite{WZ}
in the limit $x\to -1$, i.e.\ at $T<T_c$ and close to the coexistence 
curve. In this limit the function is inverted 
to give $x+1$ as a double expansion in powers of $y$ and $y^{d/2 - 1}$
\be
x+1\;=\; {\widetilde c_1} y \,+\, {\widetilde c_2} y^{d/2 - 1} \,+\,
         {\widetilde d_1} y^2 \,+\, {\widetilde d_2} y^{d/2} \,+\,
         {\widetilde d_3} y^{d-2} \,+\, \ldots \;.
\label{f_inv}
\ee
The coefficients ${\widetilde c_1}$, ${\widetilde c_2}$ and 
${\widetilde d_3}$ are then obtained from the general expression of
\cite{Brezin}. 
The above form is motivated by the $H$-dependence
in the $\epsilon$-expansion of $\chi_L$ at low temperatures \cite{WZ}.

In Fig.\ \ref{fig:eq_state} we show the function $f(x)$
from \cite{Brezin}, its low-temperature ($x\to -1$) limit and 
$f(x)$ from \cite{WZ}, which is obtained from the inverted form
of the low-temperature expression, Eq.\ (\ref{f_inv}).
We see that the low-temperature curve remains close to the
general one for a significant portion of the phase diagram, including
the critical point at $x=0$, $y=1$ and moving into the high-temperature
phase, the region to the right of the critical point.
The ``inverted'' curve and the low-temperature curve
that generated it agree quite well. (We remark that the process of 
inverting produces coefficients that are determined only to order 
$\epsilon$, while the original low-temperature expression was known 
to order $\epsilon^2$.) 
As mentioned above, the form (\ref{f_inv}) is equivalent to the 
Goldstone-singularity form for $\chi_L$ at low temperatures
if one identifies the variable $y$ with the field $H$. Nevertheless,
the fact that this form may describe the behaviour also at temperatures 
close to $T_c$ and higher is not contradictory, since the variable 
$y$ can only be identified with $H$ if $M(T,H=0)\ne 0$, 
which happens only at low temperatures.
The form (\ref{f_inv}) has explicitly nonnegative values of $y$
at $x\approx -1$, while the original perturbative
expression produces an unphysical negative $y$ in a

\setlength{\unitlength}{1cm}
\begin{picture}(13,7)
\put(-0.1,0){
   \epsfig{bbllx=127,bblly=264,bburx=451,bbury=587,
       file=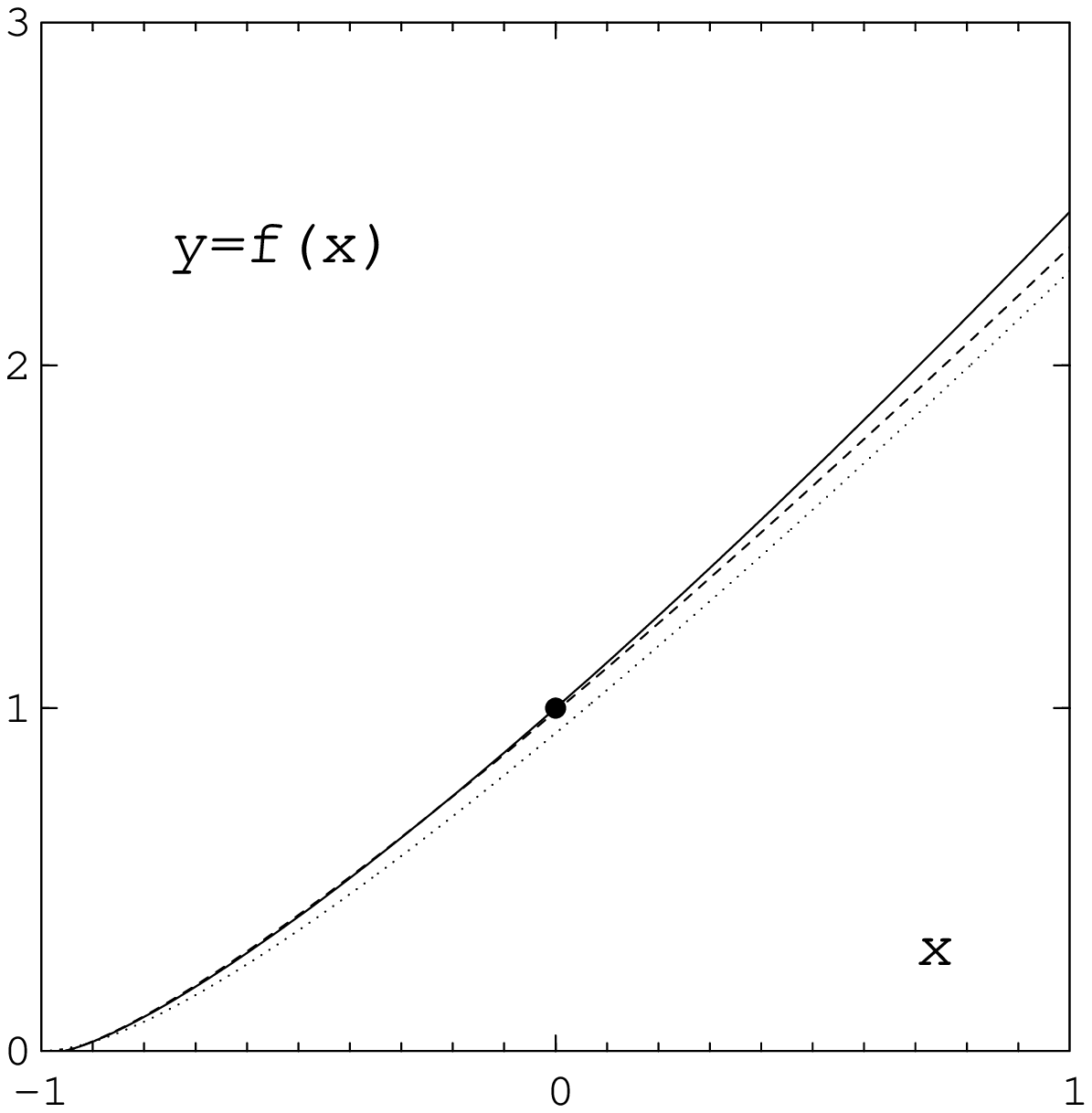, width=67mm}
          }
\put(7.5,0){
   \epsfig{bbllx=127,bblly=264,bburx=451,bbury=587,
       file=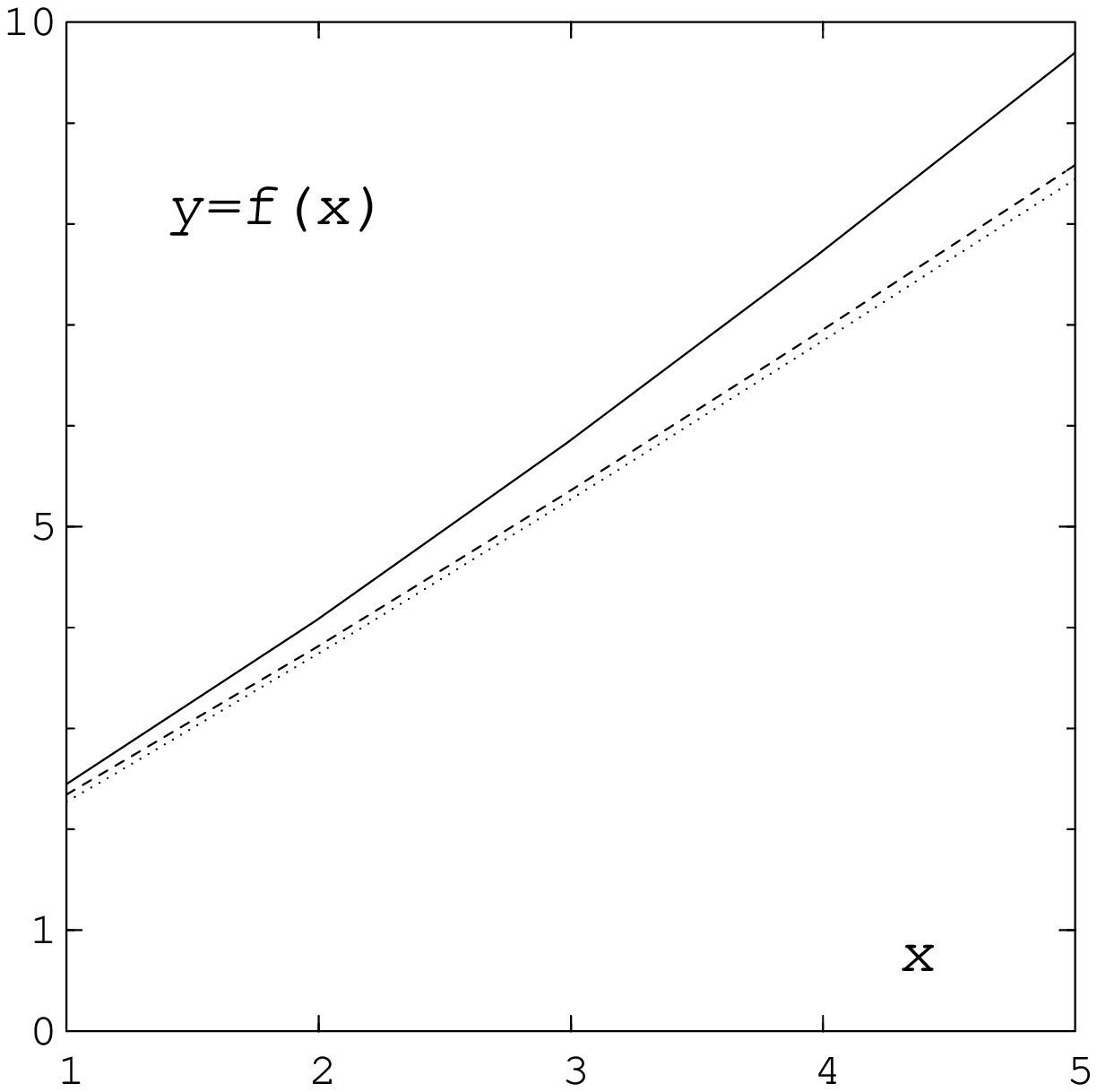, width=67mm}
          }
\end{picture}
\begin{figure}[h!]
\caption{The function $y=f(x)$ (solid line) from \cite{Brezin}, its 
approximation for $x\to -1$ (dashed line) and the inverted form (dotted line)
from \cite{WZ}.}
\label{fig:eq_state}
\end{figure}

\noindent
very small neighborhood of this point \cite{Brezin}. Possible problems with 
the form (\ref{f_inv}) are pointed out in \cite[Section 5]{pelissetto}.

As for the large-$x$ limit (corresponding to $T>T_c$ and small $H$), 
the expected behaviour is given by Griffiths's analyticity condition
\cite{Griffiths}
\be
f(x) \;=\; \sum_{n=1}^{\infty} a_n\,x^{\gamma - 2(n-1)\beta} \;.
\label{Griffiths}
\ee
None of the curves in Fig.\ \ref{fig:eq_state} approaches this limit,
since the low-temperature curves are not valid at large $x$, and
the general curve in its original form is known to have problems
in this limit \cite{Brezin}.

The perturbative equation of state has been used in \cite{RW} to
produce pictures of the expected behaviour of the $3d$ $O(4)$ model 
for a large range of temperatures and magnetic fields. 
The authors have employed an interpolation of the function from 
\cite{Brezin} with the inverted form from \cite{WZ} at low temperatures 
and the Griffiths condition at high temperatures. 
When compared to Monte Carlo data for the same model in \cite{Toussaint}, 
the perturbative scaling function shows qualitative agreement.
In Section \ref{section:sca_fun} we propose a fit of our Monte Carlo data
to the perturbative form of the equation of state, using (\ref{f_inv}).

\section{Numerical Results}
\label{section:results}

Our simulations are done on lattices with linear extensions $L=24$, 32, 48,
64, 72, 96 and 120
using the cluster algorithm of Ref.\ \cite{clust}.
We compute the magnetization $M$ and the susceptibilities $\chi_{L,T}$
at fixed $J=1/T$ (i.e.\ at fixed $T$) and varying $H$.
We note that, due to the presence of a nonzero field, the magnetization
is nonzero on finite lattices, contrary to what happens
for simulations at $H=0$, where one is led to consider 
the approximate form $ <\!1/V|\sum {\bf S}_i|\!>$.

We use the value $J_c = 0.93590$, 
obtained in simulations of the zero-field model \cite{H0}. 
In Fig.\ \ref{fig:mrhall} we show our data for the magnetization for low
temperatures up to $T_c$ plotted versus $H^{1/2}$.
We have simulated at increasingly larger values of $L$ at fixed
values of $J$ and $H$ in order to eliminate finite-size effects.
The finite-size effects for small $H$ do not disappear as one moves
away from $T_c$, but rather increase.

\n In Fig.\ \ref{fig:mroot} we plot only the results from our largest lattices.
 The solid lines are fits to the form (\ref{magn}), and the filled squares 
at $H>0$ denote the last points included in our fits. It is evident
that the predicted behaviour (linear in $H^{1/2}$) holds close to $H=0$ for 
all temperatures $T<T_c$ considered. The Goldstone-mode effects are 
therefore observable also rather close to $T_c$. 
The straight-line fits coincide with the measured points in a wide range of 
$H$ for low $T$ (for $J=1.2$ and 1.1 up to $H=0.32$).

\setlength{\unitlength}{1cm}
\begin{picture}(13,9)
\put(7.0,9){
   \epsfig{bbllx=56,bblly=255,bburx=542,bbury=583,
       file=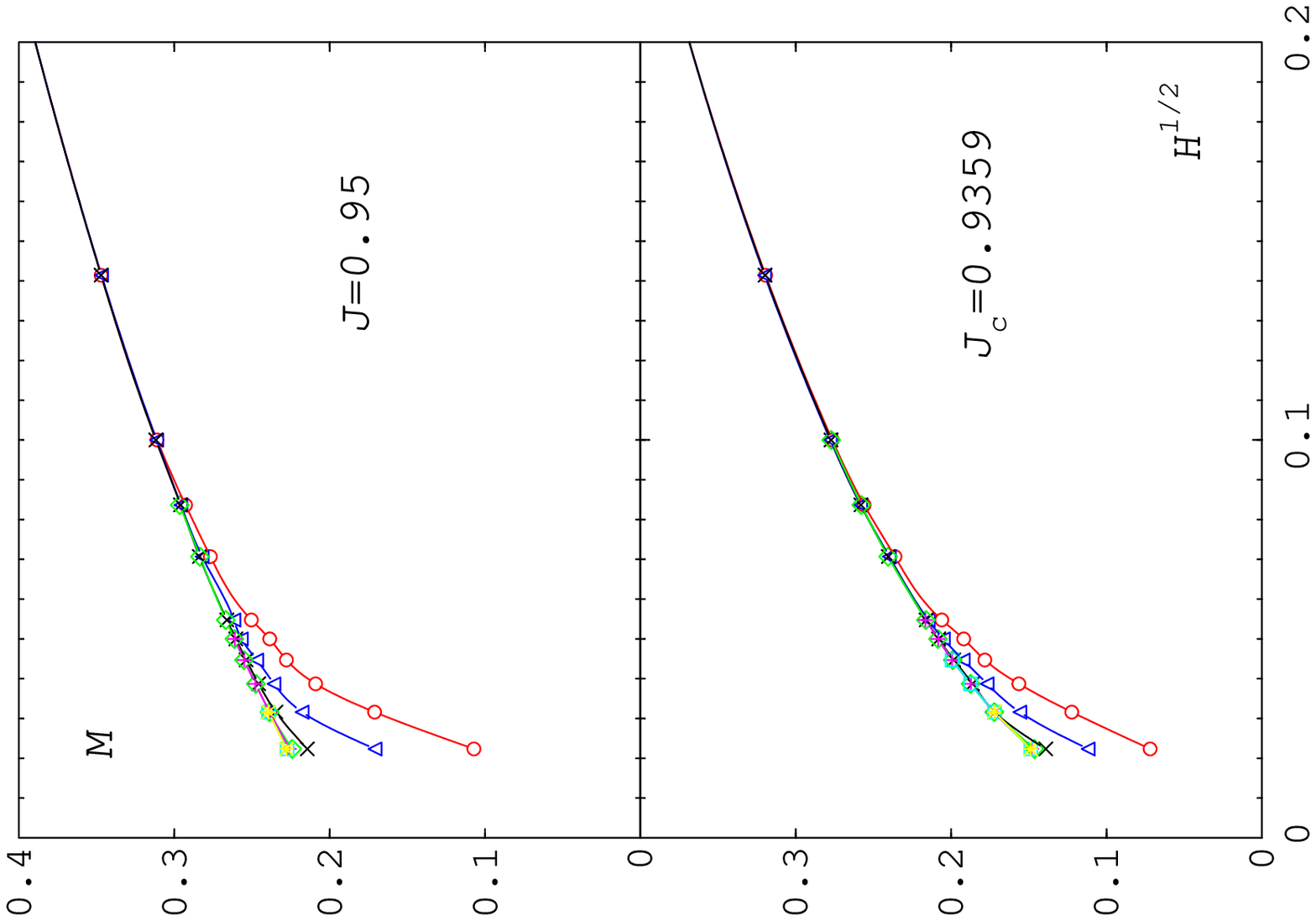, width=90mm,angle=-90}
          }
\put(0.5,9){
   \epsfig{bbllx=56,bblly=255,bburx=542,bbury=583,
       file=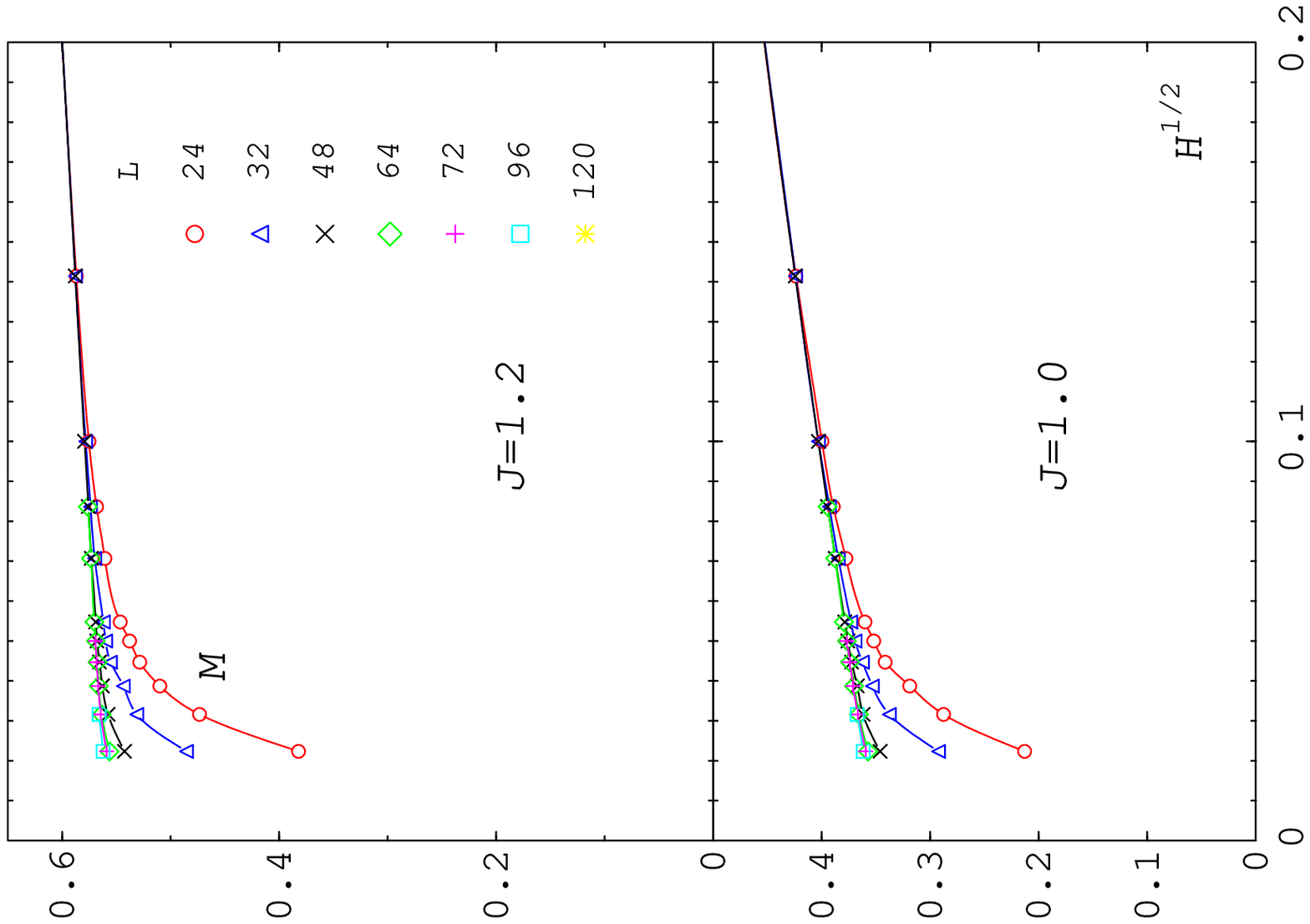, width=90mm,angle=-90}
          }
\end{picture}
\begin{figure}[h!]
\caption{The magnetization vs.\ $H^{1/2}$ in the low-temperature region
for fixed $J=1.2$, 1.0, 0.95 and $J_c$ and different lattice sizes.} 
\label{fig:mrhall}
\end{figure}

\begin{figure}[htb]
\begin{center}
   \epsfig{bbllx=127,bblly=264,bburx=451,bbury=587,
        file=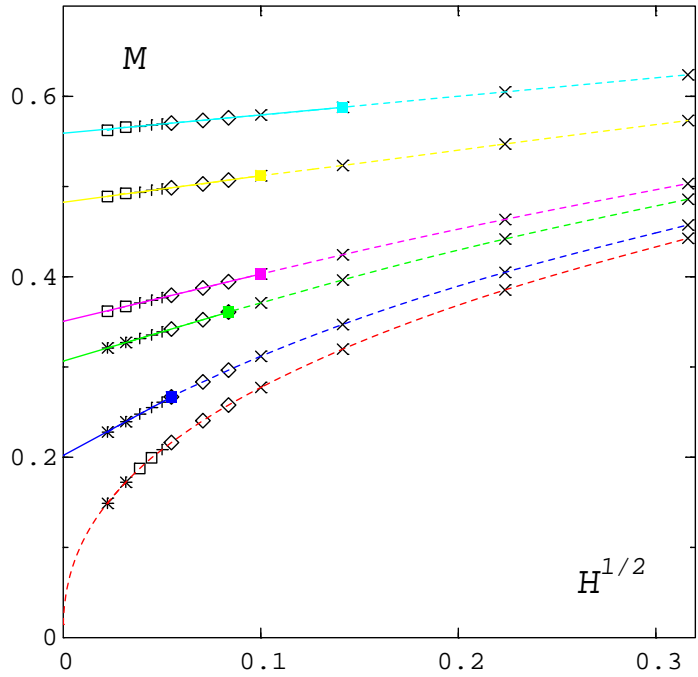,width=84mm}
\end{center}
\caption{The magnetization as a function of $H^{1/2}$ for fixed $J=1.2$,
1.1, 1.0, 0.98, 0.95 and $J_c$, starting with the highest curve.
The size $L$ is denoted as in Fig.\ \ref{fig:mrhall}.} 
\label{fig:mroot}
\end{figure}

\newpage
\begin{table}[ht]
\begin{center}
  \begin{tabular}{|c|cccc|}
    \hline
    $J=1/T$ & $M(T,0)$   & Slope $c$   & $H-$Range & $\chi^2/dof$ \\ \hline
          0.95  &0.20202(62) & 1.1825(132) &0.001-0.003  & 0.37 \\  
          0.98  &0.30650(13) & 0.6495(26)  &0.0005-0.007 & 0.38 \\ 
          1.00  &0.35061(15) & 0.5245(20)  &0.001-0.01   & 0.60 \\
          1.10  &0.48257(29) & 0.2940(38)  &0.0015-0.01  & 0.54 \\
          1.20  &0.55911(13) & 0.2033(13)  &0.002-0.02   & 0.92 \\ \hline
  \end{tabular}
\end{center}
\caption{Parameters of the fit of the magnetization to $M(T,0)+cH^{1/2}$.}

\label{tab:mfits}
\end{table}
 
\n With 
increasing $T$ the coincidence region gets smaller and vanishes at $T_c$. 
In Table \ref{tab:mfits} we have listed the fit parameters. The value $M(T,0)$
obtained from the fits is the infinite-volume value of the magnetization
on the coexistence line. In the neighbourhood of $T_c$ it should show
the usual critical behaviour
\be
M(T\ltapprox T_c,H=0) \;=\;B(T_c - T)^{\beta}  \;=\;B(1/J_c-1/J)^{\beta}~.
\ee
\n Using this simple form, without including any next-to-leading terms,
we are able to fit all points of Table \ref{tab:mfits}
with $B=0.9670(5)$ and
the exponent $\beta=0.3785(6)$, in agreement with the high-precision
zero-field determination in \cite{KK}.

As the critical point is reached the $H$-dependence of
the magnetization should change to satisfy critical scaling.
We thus fit the data from the largest lattice sizes at $T_c$ to the form
\be
M(T_c,H) \;=\;d_c H^{1/\delta}.
\label{crit}
\ee
As can be seen in Fig.\ \ref{fig:mtcab}a a very good straight-line fit 
to the largest-$L$ results is possible.
The smaller lattices show however definite finite-size effects.
We find the exponent $\delta=4.86(1)$, in agreement with \cite{KK},
and in addition we obtain the critical amplitude $d_c=0.715(1)$. 
In Fig.\ \ref{fig:mtcab}b the magnetization at $T_c$ is compared
with the finite-size-scaling prediction
\be
M(T_c,H;L) \,=\, L^{-\beta/\nu}\, Q_M(H\,L^{\beta\delta/\nu})\;,
\label{fss}
\ee
using the critical exponents of Ref.\ \cite{KK}. We observe no corrections 
to scaling, even at higher $H$-values. The scaling function $Q_M$ is 
universal. In order to be consistent with Eq.\ (\ref{crit}) for large 
$z\equiv HL^{\beta\delta/\nu}$, i.e.\ for finite small $H$ and large $L$, 
it must behave as
\be
Q_M(z)\;=\;d_cz^{1/\delta}.
\ee
\n This offers a second way to determine the critical 
amplitude $d_c$, this time exploiting also the data of the smaller lattices. 
From a fit in the $z$-range $20-1000$ we find the value $d_c=0.713(1)$,
which agrees with our first determination.

 In Fig.\ \ref{fig:clt98} we show an example (at $J=0.98$) of the different
behaviours of $\chi_T$ and $\chi_L$ at low temperatures. As a test we compare
the result for $\partial M/\partial H$ (line) from the $M$-fits in 
Table\ \ref{tab:mfits} to the $\chi_L$-data. Though there are large finite-size 
effects for  small $L$, the results for the highest $L$-values agree nicely 
with the expected behaviour. A similar test can be done for
$\chi_T$ by showing also the result for $M/H$ (line), as obtained from the 
measurements of the magnetization in Fig.\ \ref{fig:mroot}. Here as well
we find agreement for large $L$. 


\setlength{\unitlength}{1cm}
\begin{picture}(13,7)
\put(-0.1,0){
   \epsfig{bbllx=127,bblly=264,bburx=451,bbury=587,
       file=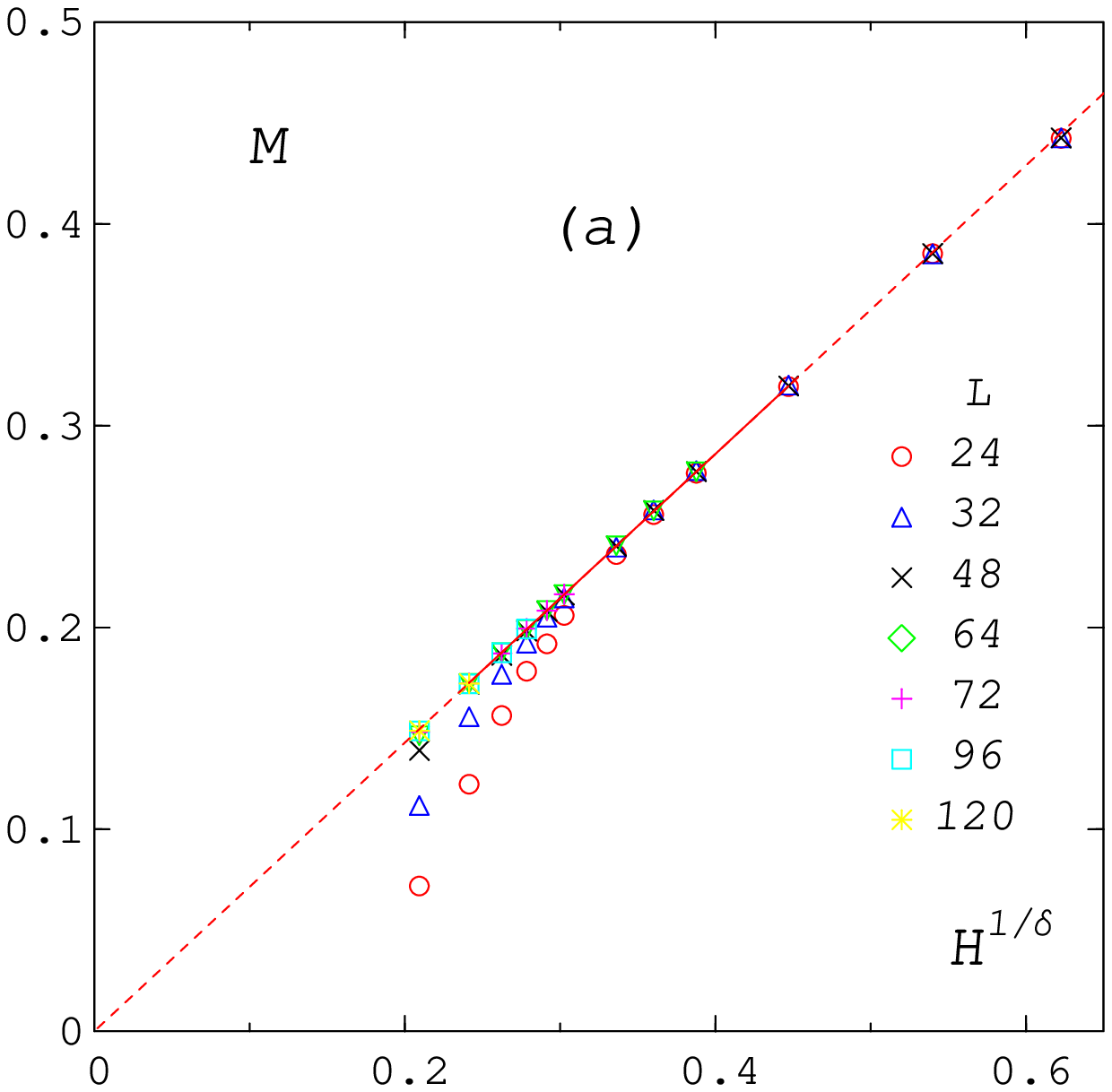, width=67mm}
          }
\put(7.5,0){
   \epsfig{bbllx=127,bblly=264,bburx=451,bbury=587,
       file=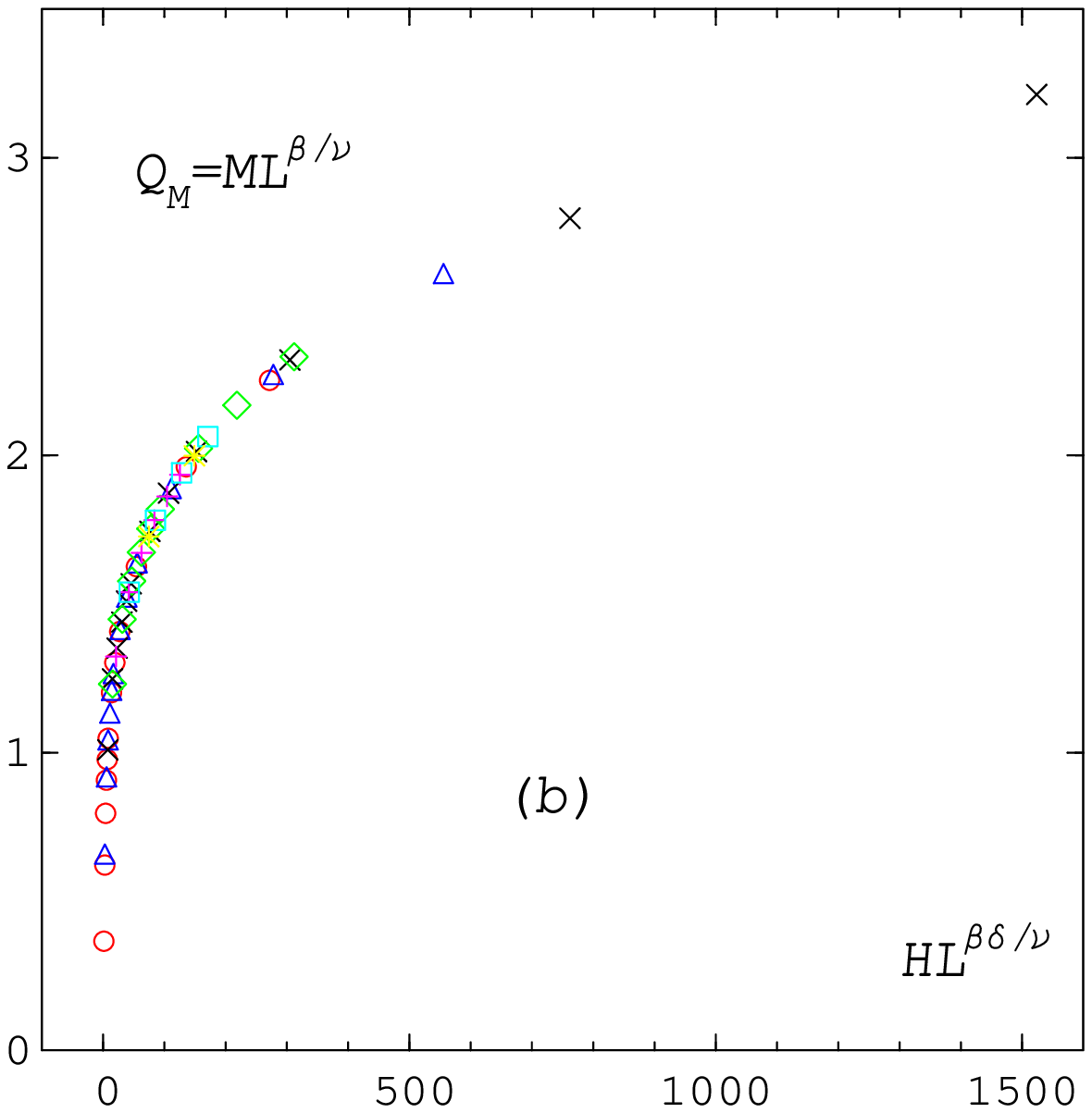, width=67mm}
          }
\end{picture}
\begin{figure}[h!]
\caption{The magnetization at $T_c$. In the left figure (a), $M$ is plotted 
vs.\ $H^{1/\delta}$, the line is the fit\ (\ref{crit}), the solid part 
shows the range used for the fit. The right plot (b) shows the 
finite-size-scaling function $Q_M$ from Eq.\ (\ref{fss}).}
\label{fig:mtcab}
\end{figure}


\setlength{\unitlength}{1cm}
\begin{picture}(13,7)
\put(-0.1,0){
   \epsfig{bbllx=127,bblly=264,bburx=451,bbury=587,
       file=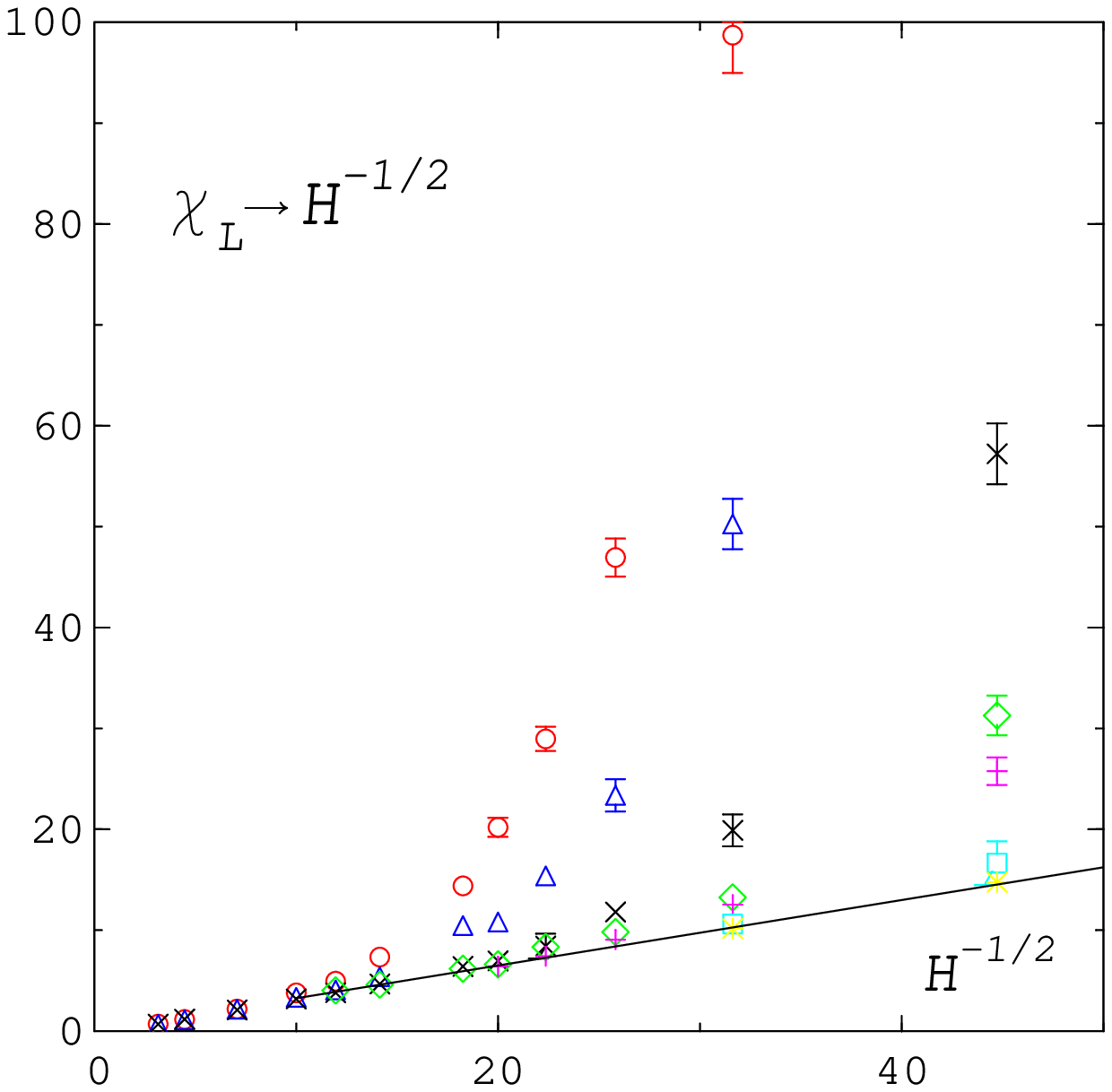, width=67mm}
          }
\put(7.5,0){
   \epsfig{bbllx=127,bblly=264,bburx=451,bbury=587,
       file=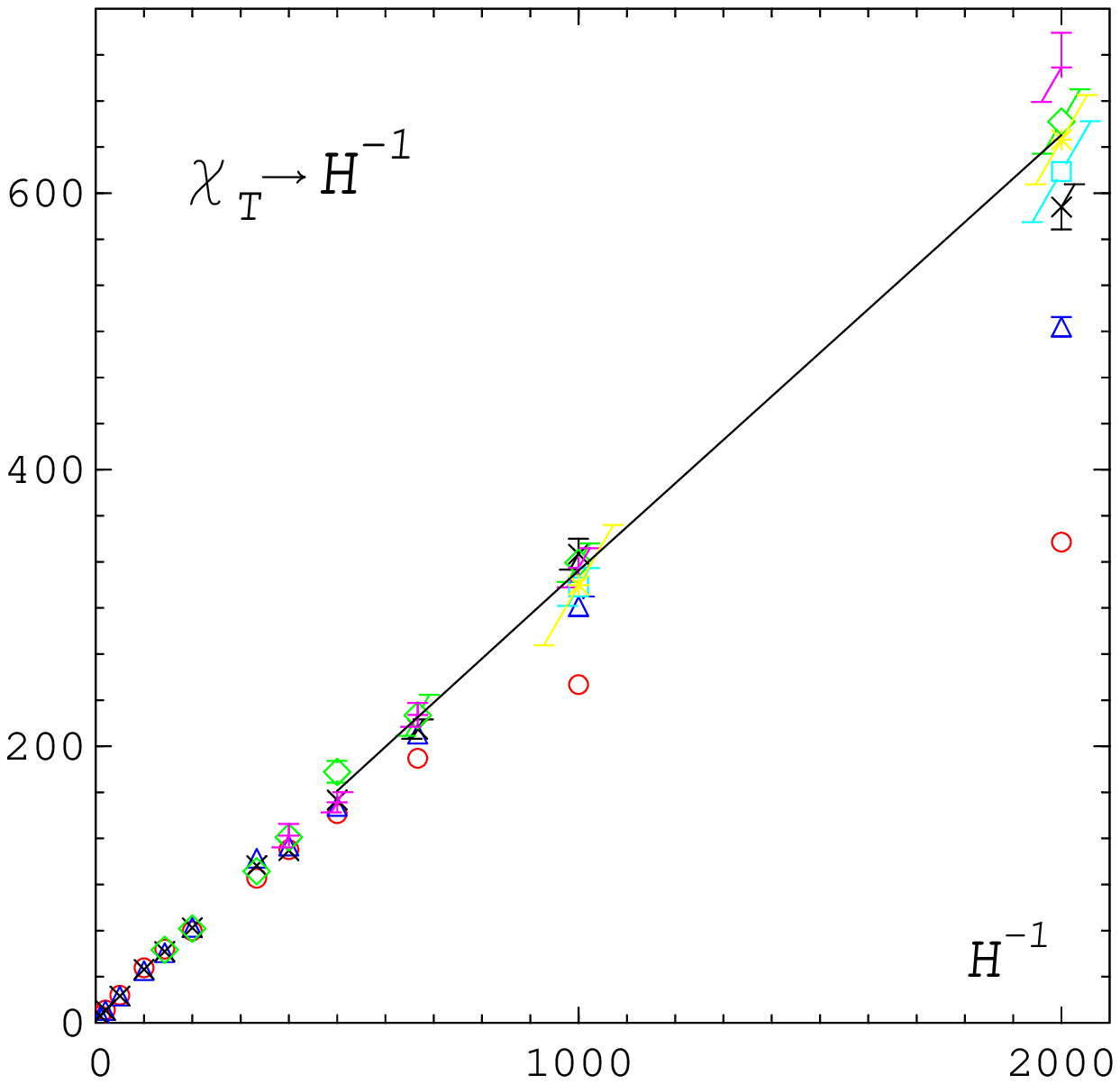, width=67mm}
          }
\end{picture}
\begin{figure}[h!]
\caption{The susceptibilities below $T_c$ at $J=0.98$. On the left $\chi_L$ 
is plotted vs.\ $H^{-1/2}$. The right plot shows $\chi_T$ as a function 
of $H^{-1}$. The lines are explained in the text.}
\label{fig:clt98}
\end{figure}

The $H$-dependencies of the susceptibilities at the critical point are the 
same. Their amplitudes differ however by a factor $\delta$. Here and in the
following we use the values $\delta=4.86$ and $\beta=0.38$~.
From Eqs.\ (\ref{chil}) and (\ref{WI}) and from the magnetization at 
the critical point, Eq.\ (\ref{crit}), we derive
\be
\chi_L\;=\;(d_c/\delta) H^{1/\delta -1}~\quad {\rm and} \quad 
\chi_T\;=\;d_c H^{1/\delta -1}~.
\label{chiatc}
\ee
In the left part of Fig.\ \ref{fig:cltrest} we compare $\chi_L$ and $\chi_T$
 at $T_c$. The lines in the figure are calculated from Eq.\ (\ref{chiatc}) and
the fit results of Eq.\ (\ref{crit}). We see again consistency with the 
highest-$L$ data. The right part of Fig.\ \ref{fig:cltrest} shows two examples 
of $\chi_L$ and $\chi_T$ for high temperatures. 
Both susceptibilities converge to one value $\chi(T)$ for $H\to 0$, since no  
spontaneous symmetry breaking occurs for $T>T_c$. At the higher temperature 
corresponding to $J=0.80$ the two susceptibilities are essentially constant
(i.e.\ the magnetization is linear in $H$) and equal for a large range 
in $H$. However, at $J=0.90$ (that is closer to $T_c$) and finite $H$-values
$\chi_T$ is larger than $\chi_L$. 

\setlength{\unitlength}{1cm}
\begin{picture}(13,7)
\put(-0.1,0){
   \epsfig{bbllx=127,bblly=264,bburx=451,bbury=587,
       file=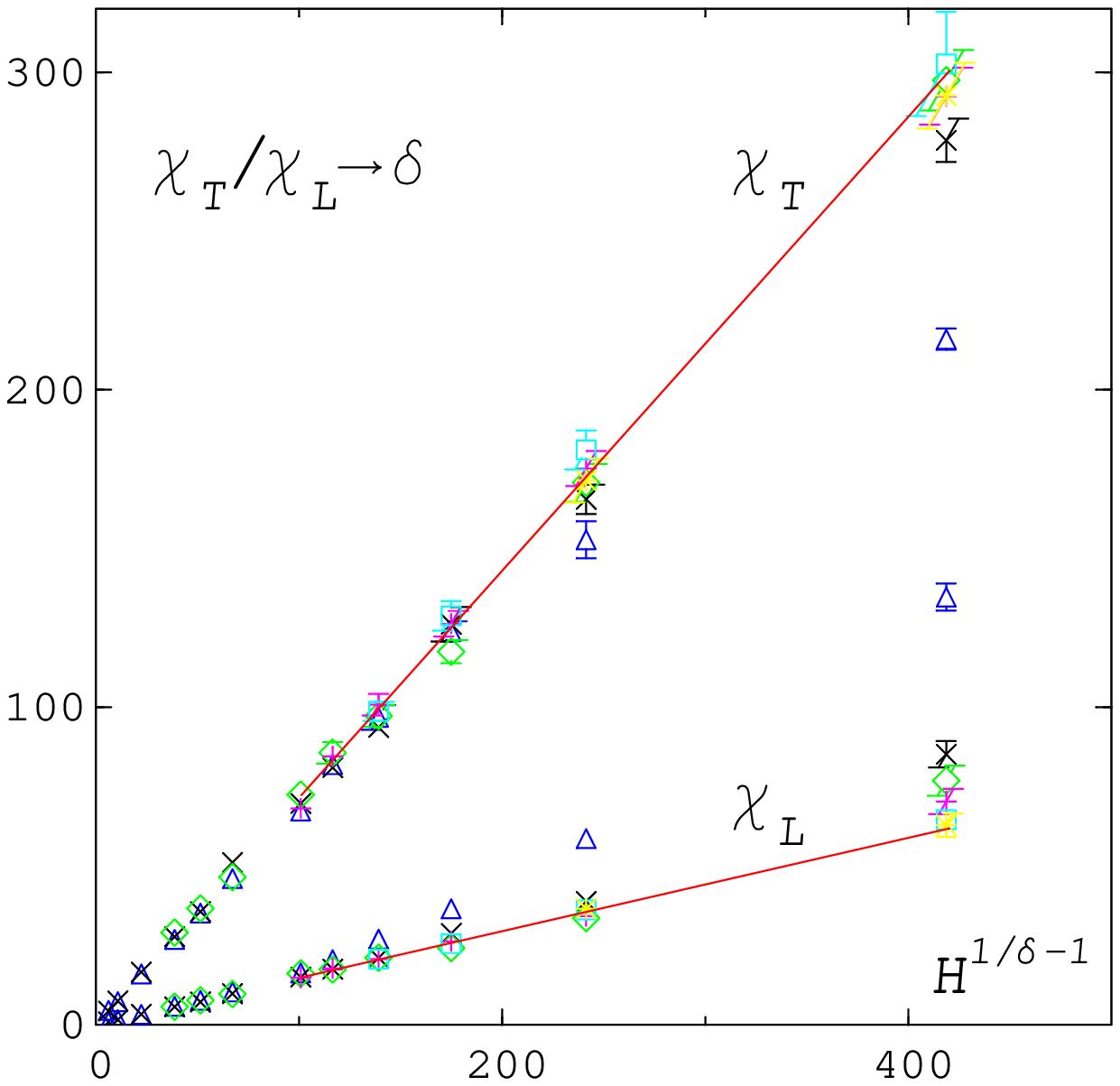, width=67mm}
          }
\put(7.5,0){
   \epsfig{bbllx=127,bblly=264,bburx=451,bbury=587,
       file=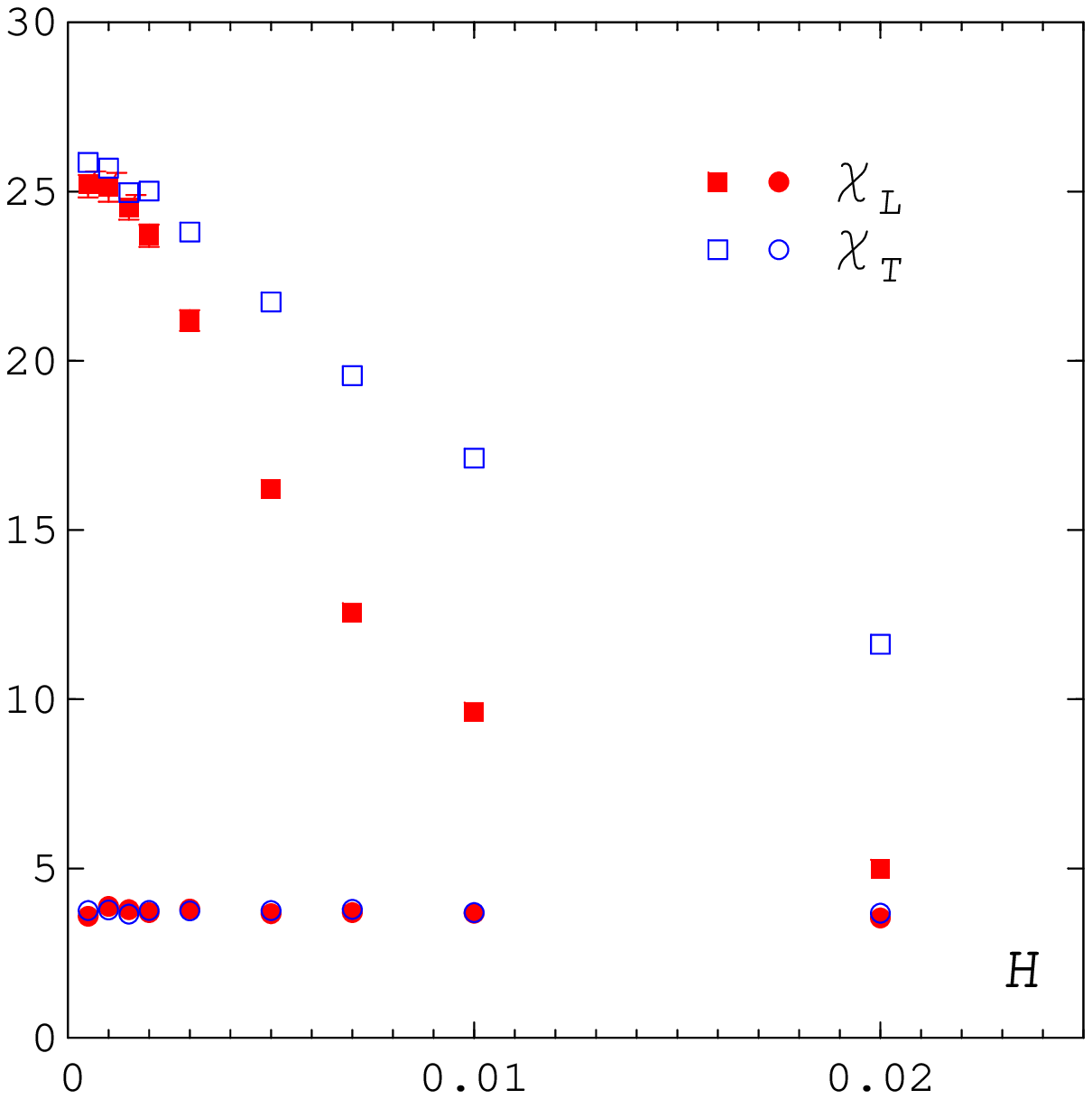, width=67mm}
          }
\end{picture}
\begin{figure}[h!]
\caption{The susceptibilities at $T_c$ (left) vs.\ $H^{1/\delta -1}$ and above 
$T_c$ (right) vs.\ $H$ at $J=0.90$ (squares) and $J=0.80$ (circles).
The lines are explained in the text.}
\label{fig:cltrest}
\end{figure}
\section{The Scaling Function}
\label{section:sca_fun}


The scaling function for the three-dimensional $O(4)$ model
was determined from a fit of Monte Carlo data in \cite{Toussaint}.
Our goal here is to describe our data using the perturbative 
form of the equation of state as discussed in Section \ref{section:PT},
but with nonperturbative coefficients, determined from a fit of the data.
The original form of the function $y=f(x)$ from \cite{Brezin} is not 
suitable for such a fit. We thus consider Eq.\ (\ref{f_inv}), which is 
written as a simple series expansion in $y$. We do not expect this
form to describe the data for all $x$ and $y$, yet looking at
Fig.\ \ref{fig:eq_state}, we hope to cover a significant
portion of the phase diagram for small $x$. The idea is to interpolate
this result with a fit to the large-$x$ form (\ref{Griffiths}).

\begin{figure}[htb]
\begin{center}
   \epsfig{bbllx=127,bblly=264,bburx=451,bbury=587,
        file=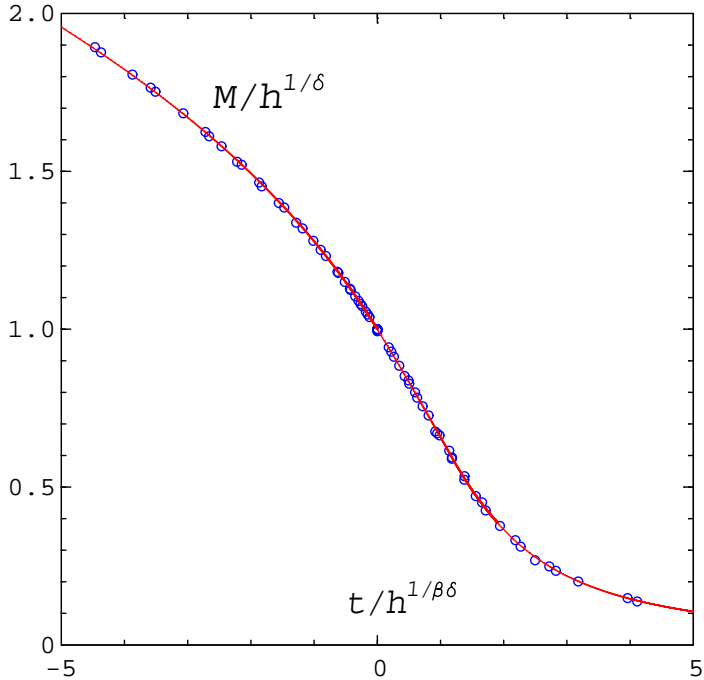,width=84mm}
\end{center}
\caption{The scaling function $f_G=M/h^{1/\delta}$ from Eq.\ (\ref{ftous}). 
The data are shown together with our fit from Eq.\ (\ref{totalfit}).}
\label{fig:scafun}
\end{figure}

Our fits are shown, together with our data, in Figs.\ 
\ref{fig:yfx} and \ref{fig:scafun}.
We have considered data from our largest lattices, for inverse
temperatures $0.9 \le J \le 1.0$ and magnetic fields $H \leq 0.01$.
The normalization constants $H_0$ and $T_0$, obtained from Eq.\ (\ref{normal})
and our fits in Section \ref{section:results} are given by
\be
H_0\;=\; 5.08(3)\;, \quad  T_0\;=\;1.093(2)\;.
\ee
We have performed a fit using the three leading terms
in (\ref{f_inv}) for small $y$ 
\be
x_1(y)+1 \;=\; ({\widetilde c_1} \,+\, {\widetilde d_3})\,y \,+\,
             {\widetilde c_2}\,y^{1/2} \,+\, 
             {\widetilde d_2}\,y^{3/2} \;.
\label{PTform}
\ee
This form was fitted in the interval $-1<x\ltapprox 1.5$, giving
\be
{\widetilde c_1} + {\widetilde d_3} \,=\, 0.345(12) \;,\quad
{\widetilde c_2} \,=\, 0.6744(73) \;,\quad
{\widetilde d_2} \,=\, -0.0232(49) \;.
\ee
\setlength{\unitlength}{1cm}
\begin{picture}(13,7)
\put(-0.1,0){
   \epsfig{bbllx=127,bblly=264,bburx=451,bbury=587,
       file=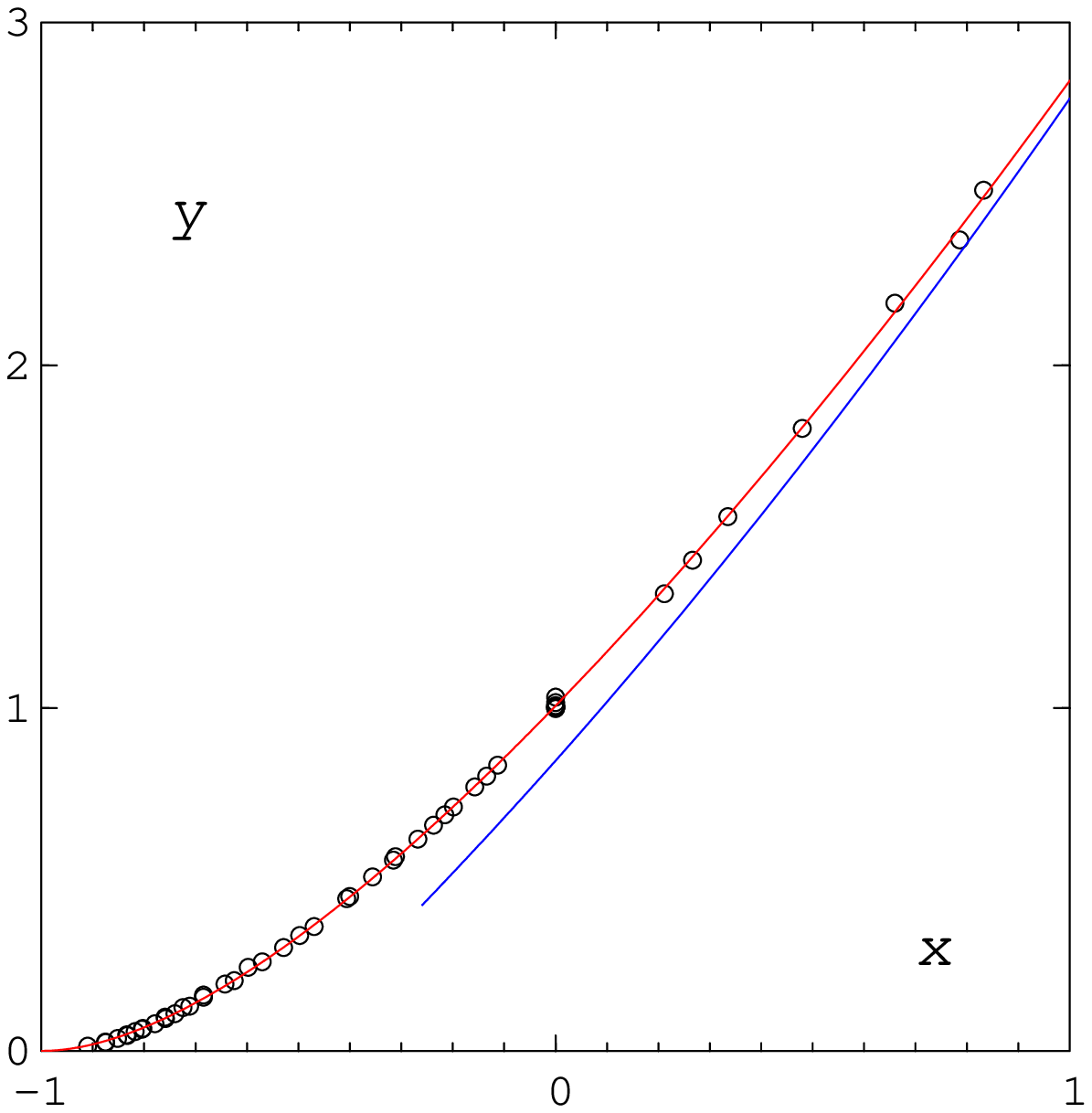, width=67mm}
          }
\put(7.5,0){
   \epsfig{bbllx=127,bblly=264,bburx=451,bbury=587,
       file=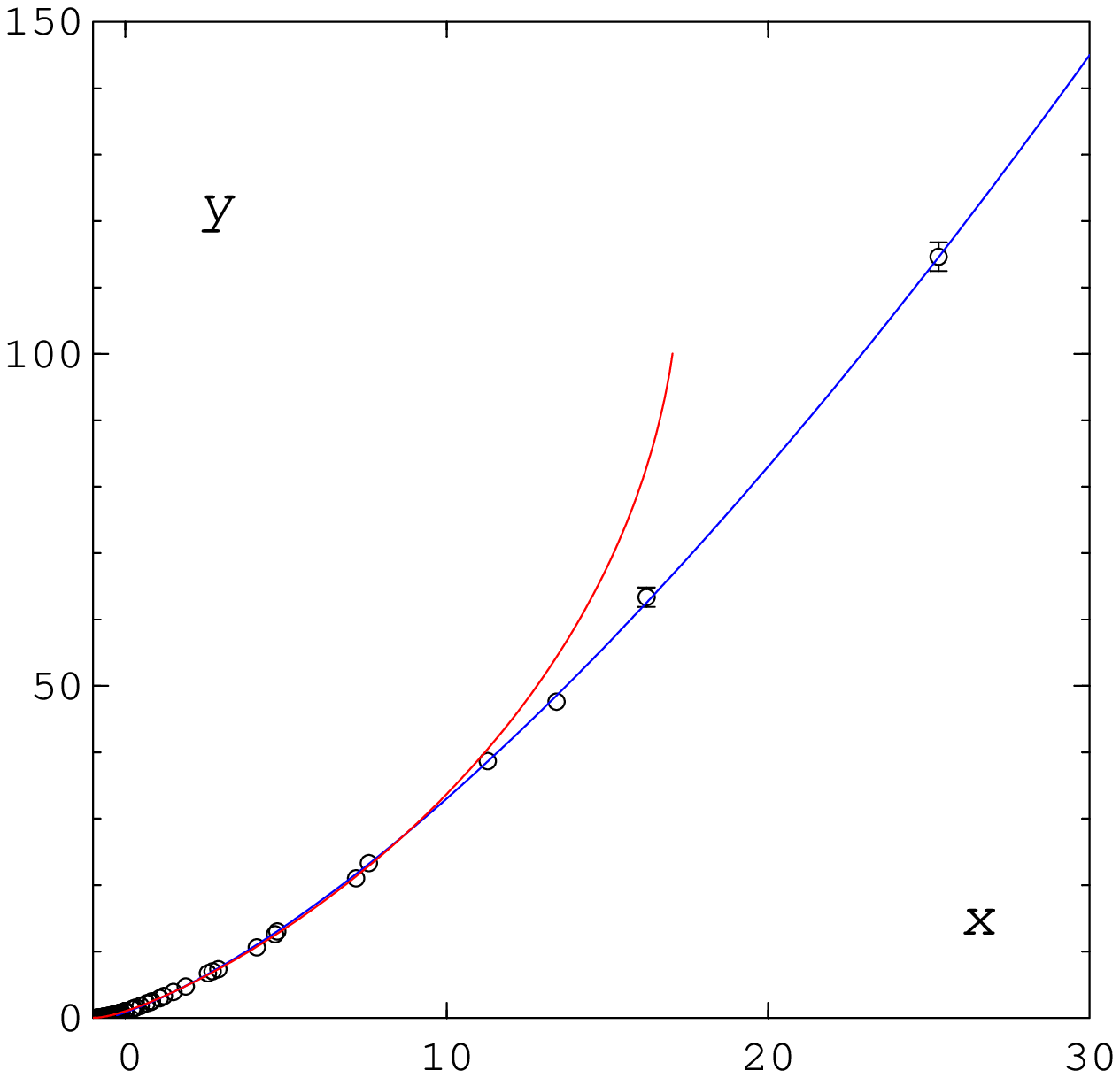, width=67mm}
          }
\end{picture}
\begin{figure}[h!]
\caption{The function $y=f(x)$ from fits to data at small $x$ (red line) 
and at large $x$ (blue line).}
\label{fig:yfx}
\end{figure}

\n The fit describes all the data at $T<T_c$ and also higher, up to
$x\approx 5$. This confirms that the expression (\ref{f_inv}) is
valid also away from $x\approx -1$, as observed in Section
\ref{section:PT}. We note the small value of ${\widetilde d_2}$.
An attempt to include the next power of $y$ leads to a coefficient
that is zero within errors.
We also see that our data are not sensitive to possible logarithmic
corrections to Eq.\ (\ref{f_inv}) as proposed in \cite{pelissetto}.
Our coefficients can be compared to those calculated perturbatively for
$N=4$ in Ref.\ \cite{WZ}
\be
{\widetilde c_1} + {\widetilde d_3} \,=\, 0.528 \;,\quad
{\widetilde c_2} \,=\, 0.530 \;.
\ee

For large $x$ we have done a 2-parameter fit
of the behaviour (\ref{Griffiths}), in the corresponding form for
$x$ in terms of $y$
\be
x_2(y)\;=\; a\, y^{1/\gamma} \,+\, b\,y^{(1-2\beta)/\gamma}~.
\label{highx}
\ee
Considering data points with $y>50$ (corresponding to 
$x\gtapprox 15$) we obtain
\be
a \,=\, 1.084(6)\;, \quad  b \,=\, -0.874(25) \;.
\ee
Expression (\ref{highx}) is seen to describe the data for
$x \gtapprox 2$. We mention that a fit of our data to the
leading term in Griffiths's condition, using $x\geq 50$,
yields $\gamma = 1.45(1)$, which is in agreement with \cite{KK}.

The small- and large-$x$ curves cover the whole range
of values of $x$ remarkably well. In fact, the two curves are 
approximately superimposed in the interval $2\ltapprox x \ltapprox 8$. 
We can therefore interpolate smoothly, for example by taking
\be
x(y) \;=\; x_1(y)\,\frac{y_0^3}{y_0^3 + y^3} \,+\,
           x_2(y)\,\frac{y^3}{y_0^3 + y^3}
\label{totalfit}
\ee
at $y_0=10$, which corresponds to $x\approx 4$.
Expression (\ref{totalfit}) is equivalent to the equation of
state (\ref{eqstate}) and to the scaling function $f_G$ in
(\ref{ftous}).
In Fig.\ \ref{fig:scafun} we show a plot of $f_G$ obtained
parametrically from $x(y)$ in (\ref{totalfit}).
The two variables in the plot are simply related to $x$ and $y$ by
\be
f_G \;=\; M/h^{1/\delta} \;=\; y^{-1/\delta}\;, \quad  
t/h^{1/\beta\delta} \;=\; x\,y^{-1/\beta\delta} \;.
\ee
We see a remarkable agreement of our data points with the form
suggested by perturbation theory. 
With respect to the scaling function in \cite{Toussaint} our function
is slightly higher for large negative $t/h^{1/\delta\beta}$, due to
our more complete elimination of finite-size effects.


\section{Summary and Conclusions}
\label{section:conclusion}

We have shown that the Golstone singularities are clearly observable
at low temperatures, and also close to $T_c$.
In fact, we are able to {\em use} the observed Goldstone-effect behaviour to 
extrapolate our data to $H\to0$ and obtain the zero-field critical exponent
$\beta$ in good agreement with \cite{KK}. 
We remark that the same does {\em not} happen at high temperatures: we
are not able to get the exponent $\gamma$ from extrapolations using the constant 
behaviour of the longitudinal susceptibility 
(or the linear behaviour of the magnetization), since this behaviour is masked
close to $T_c$ for the fields $H$ we have taken into account.
At the same $H$'s the $H^{1/2}$ behaviour is clearly present
for all the $T<T_c$ we consider, showing that the Goldstone effect 
is dominating the critical one, except at $T_c$.

A strong manifestation of the Goldstone behaviour had been conjectured
perturbatively \cite{WZ}, based on the size of the coefficient
${\widetilde c_2}$ in the $\epsilon$-expansion of the
equation of state. We have fitted the perturbative form in Section
\ref{section:sca_fun}, finding a coefficient that is even larger
than the perturbative one.

The resulting curve for the equation of state describes all the data
beautifully, and can be plotted parametrically for the scaling function.

As a by-product of our work we have determined 
the critical exponent $\delta=4.86(1)$
by a fit of the magnetization at $T_c$ to the critical scaling behaviour
as a function of $H$. In addition we checked the finite-size-scaling 
prediction for $M$. It is remarkable that we observed in both cases no 
corrections to scaling. 

A similar investigation for the $O(2)$ model is currently being done
\cite{inprep}. 
\vskip 0.5truecm
\noindent{\Large{\bf Acknowledgements}}
\bigskip

We thank Attilio Cucchieri and Frithjof Karsch for helpful suggestions and
comments.
Our computer code for the cluster algorithm was based on a zero-field
program by Manfred Oevers.
This work was supported by the Deutsche Forschungs\-ge\-meinschaft
under Grant No.\ Ka 1198/4-1.

\clearpage

\end{document}